\documentclass[11pt]{article}
\usepackage[utf8]{inputenc}
\usepackage{amsmath,amsfonts,mathtools,dsfont,booktabs,tikz,float,pgfplots,subcaption,wrapfig,esint,enumitem,siunitx,authblk,amssymb}
\usepackage[colorlinks=true,linkcolor=blue,citecolor=blue,urlcolor=blue,bookmarks=true]{hyperref}
\usetikzlibrary{calc}
\graphicspath{{./graphics/}}
\usepackage[final]{pdfpages}
\usepackage[labelfont=bf]{caption}

\hypersetup{
    colorlinks,
    linkcolor={blue!50!black},
    citecolor={blue!50!black},
    urlcolor={blue!80!black}
}
\usepackage[capitalize,noabbrev,nameinlink]{cleveref}

\usepackage{physics}
\usepackage{textcomp}
\usepackage{geometry}[margin=1in]
\usepackage{tensor}

\usepackage{nicefrac}

\newcommand{\GF}[1]{\mathbb{F}_{#1}}

\usepackage{amsthm}
\newtheorem{theorem}{Theorem}

\newtheorem{corollary}{Corollary}

\theoremstyle{definition}
\newtheorem{definition}{Definition}

\usepackage[square,numbers,sort&compress]{natbib}
\title{Algebraic Geometry Codes and Decoded Quantum Interferometry}
\author[1,2]{Andi Gu}
\author[1]{Stephen P. Jordan}
\affil[1]{\small{Google Quantum AI}}
\affil[2]{\small{Department of Physics, Harvard University, Cambridge, MA 02138, USA}}

\date{}

\begin{document}

\maketitle

\begin{abstract}
Decoded Quantum Interferometry (DQI) defines a duality that pairs decoding problems with optimization problems. The original work on DQI considered Reed-Solomon decoding, whose dual optimization problem, called Optimal Polynomial Intersection (OPI), is a polynomial regression problem over a finite field. Here, we consider a class of algebraic geometry codes called Hermitian codes, which achieve block length $q^3$ using alphabet $\GF{q^2}$ compared to Reed-Solomon's limitation to block length $q$ over $\GF{q}$, requiring approximately one-third fewer qubits per field element for quantum implementations. We show that the dual optimization problem, which we call Hermitian Optimal Polynomial Intersection (HOPI), is a polynomial regression problem over a Hermitian curve, and because the dual to a Hermitian code is another Hermitian code, the HOPI problem can also be viewed as approximate list recovery for Hermitian codes. By comparing to Prange's algorithm, simulated annealing, and algebraic list recovery algorithms, we find a large parameter regime in which DQI efficiently achieves a better approximation than these classical algorithms, suggesting that the apparent quantum speedup offered by DQI extends beyond Reed-Solomon codes to a broader class of polynomial regression problems on algebraic varieties.
\end{abstract}

\section{Introduction}

Combinatorial optimization is integral to a wide spectrum of scientific and engineering tasks, yet many problem instances are conjectured to resist efficient classical solution. Quantum algorithms offer potential speed‑ups, but rigorous evidence for substantial advantages remains scarce. \emph{Decoded Quantum Interferometry} (DQI) was introduced as a framework that turns efficient \emph{classical} error‑correcting‑code decoders into \emph{quantum} optimization primitives. By applying a decoder coherently to a superposition of error strings, DQI efficiently prepares solutions in time polynomial in the problem size~\cite{dqi}.

The original work realized this idea with Reed–Solomon (RS) codes and found a class of constraint‑satisfaction problems for which DQI efficiently finds better approximate optima than known classical algorithms. The dual optimization problem, called Optimal Polynomial Intersection (OPI), is a polynomial regression problem over a finite field where the goal is to find a univariate polynomial with bounded degree that maximizes agreement with given constraint sets at specified field elements.

For the OPI problem, DQI with Reed-Solomon codes demonstrated an apparent exponential quantum speedup for polynomial regression over finite fields. While this initial result established the power of the DQI framework, it naturally raises the question of whether the apparent exponential speedup extends beyond Reed-Solomon codes to other families of algebraic codes. Reed-Solomon codes, though fundamental and well-studied, are just one instance of a much richer family of codes, known as algebraic geometry codes. Other codes within this family offer superior parameters or structural properties that could be advantageous in the context of DQI (see~\cref{app:ag-code} for background on algebraic geometry codes).

Indeed, algebraic geometry codes have found significant applications in quantum error correction and magic state distillation. Most notably,~\cite{krishna2019towards} used Reed-Solomon codes to construct triorthogonal quantum codes enabling magic state distillation with asymptotically vanishing overhead ($\gamma \to 0$), exploiting polynomial evaluation's algebraic structure for transversal non-Clifford gates. Recent breakthroughs have extended this further, with~\cite{wills2024constant} achieving constant-overhead magic state distillation using more sophisticated AG constructions. These developments demonstrate that the same algebraic structure enabling efficient classical decoding in AG codes—which we leverage in DQI—also provides the geometric framework for fault-tolerant quantum protocols~\cite{campbell2012magic,golowich2024asymptotically}.

Hermitian codes present an ideal candidate for extending the DQI framework. Constructed from the Hermitian curve $y^q + y = x^{q+1}$ over $\GF{q^2}$, these codes achieve block length $q^3$ using alphabet $\GF{q^2}$ (see \cref{app:hermitian}), compared to Reed-Solomon codes, which are limited to block length $q$ over alphabet $\GF{q}$. This represents a fundamental advantage for quantum implementations: while Reed-Solomon codes require $\lceil\log_2(n)\rceil$ qubits per field element for block length $n$, Hermitian codes achieve the same block length using only $\lceil(2/3)\log_2(n)\rceil$ qubits per element. This decrease in quantum resource requirements becomes increasingly significant as problem sizes grow. These codes achieve excellent parameters (large length over a relatively small field, strong rate–distance trade-offs) and are well understood~\cite{stichtenoth1993}. Crucially for DQI applications, Hermitian codes admit efficient classical decoding algorithms that can correct errors up to half of the minimum distance, providing the necessary foundation for the syndrome decoding step in DQI.

In this work, we demonstrate that the DQI framework generalizes naturally to Hermitian codes and efficiently finds better approximate optima than Prange's algorithm. This generalizes DQI to apply to problems in algebraic geometry. In the HOPI problem, we optimize a polynomial function to satisfy constraints regarding its values on an algebraic variety, specifically the Hermitian curve (\cref{eq:hermite}). The OPI problem studied in \cite{dqi} is the special case where the algebraic variety is simply the finite field $\GF{p}$ and the fitting function is restricted to polynomials rather than rational functions. In \cite{dqi} the multivariate OPI problem is also studied, which generalizes OPI, but still treats the problem of fitting on the entire affine space over which polynomial in $h$ variables is defined, namely $\GF{P}^h$. Here, we extend DQI beyond affine spaces to solve optimization problems on nontrivial algebraic curves.

We formulate the Hermitian Optimal Polynomial Intersection problem as a direct generalization of the original OPI problem and, via the semicircle law of \cite{dqi}, establish quantum advantage over the classical algorithms that we believe to be the most relevant competitors: Prange's algorithm, algebraic list decoding methods, and simulated annealing. Our results suggest that the quantum advantage observed in DQI is not specific to Reed-Solomon codes but rather reflects a fundamental property of the algorithmic framework when applied to well-structured algebraic codes.

\section{Related Work}

Our work builds on several strands of prior research that connect quantum algorithms, algebraic geometry codes, and classical decoding. Decoded Quantum Interferometry (DQI) was introduced by~\cite{dqi}, which proved the general semicircle law and instantiated the framework with Reed–Solomon and Reed–Muller codes. DQI itself builds on earlier Fourier-transform-based reductions for lattice and coding problems \cite{ATS03,R04,AR05,regev2009}, and is conceptually related to the ``filtering'' algorithm~\cite{chen2021quantumalgorithmsvariantsaveragecase}, which achieved exponential speedups relative to known classical algorithms for lattice problems via intrinsically quantum decoding. More recent developments in this line include the query-complexity separation for a folded Reed–Solomon variant of max-LINSAT~\cite{Yamakawa_2024}, novel quantum decoding methods based on unambiguous state discrimination~\cite{chailloux2023quantumdecodingproblem}, and work showing the performance of DQI can be improved by decoders that take advantage of ``soft'' information~\cite{CT24}. Our extension of DQI to Hermitian codes situates it within this broader exploration of how Regev's reduction~\cite{regev2009} can be leveraged for quantum advantage.

Within coding theory, algebraic geometry codes, introduced in Ref.~\cite{goppa1983} (see \cref{app:ag-code}), have provided powerful constructions that go beyond Reed–Solomon. The Hermitian curve $y^q + y = x^{q+1}$ is particularly important (see \cref{app:hermitian}), giving rise to codes of length $q^3$ over $\GF{q^2}$ with good distance and rate properties.\footnote{A classical reference for this is Stichtenoth's treatment of algebraic function fields and codes~\cite{stichtenoth1993}.} Efficient decoding algorithms for Hermitian codes have been studied extensively, beginning with syndrome-based approaches and the Feng–Rao algorithm~\cite{fengrao}, and extending to list-decoding methods inspired by the Guruswami–Sudan framework~\cite{gs1998}. A feature especially relevant for DQI is the duality property: the dual of a Hermitian code is again a Hermitian code with comparable parameters, ensuring that both the code and its dual admit efficient decoders.

From the perspective of computational complexity, syndrome decoding—the problem of finding an error vector given its syndrome (a linear transformation of the error)—is NP-hard in general~\cite{berlekamp1978}, which motivates the study of efficient decoding algorithms for structured code families. For Reed–Solomon codes, different decoding algorithms handle different error regimes. \emph{Unique decoding} algorithms guarantee to find the unique closest codeword when the number of errors is small enough. The Berlekamp–Massey algorithm~\cite{massey,berlekamp2015} provides polynomial-time unique decoding up to $\lfloor (d-1)/2 \rfloor$ errors, where $d$ is the minimum distance (the smallest Hamming distance between distinct codewords). \emph{List decoding} algorithms can handle more errors by returning a small list of candidate codewords rather than a single answer. The Guruswami–Sudan algorithm~\cite{gs1998} extends the error-correcting radius beyond the unique decoding bound, specifically up to $d(1 - \sqrt{R})$ errors for rate $R$ codes, while returning a polynomial-size list of possible codewords. \emph{List recovery} is a generalization where, instead of a single noisy codeword, we are given ``lists'' of possible values at each coordinate and must find codewords that are consistent with these constraint sets. Analogous list-recovery algorithms exist for Hermitian and other algebraic geometry codes, although their practical guarantees typically require the input lists to be small relative to the alphabet size (typically, up to polylogarithmic in alphabet size) --- a limitation that becomes restrictive when list sizes approach the field size. In the present work, we follow~\cite{dqi} in comparing against Prange's information set decoding algorithm~\cite{prange}, which serves as the canonical classical baseline for the syndrome decoding problem in general.

Finally, our formulation of the Hermitian Optimal Polynomial Intersection (HOPI) problem extends the Optimal Polynomial Intersection (OPI) problem studied in~\cite{dqi}, which can itself be viewed as a special case of list-recovery. This perspective originates in classical work on list decoding for Reed–Solomon and algebraic geometry codes, most prominently the Guruswami–Sudan algorithm~\cite{gs1998}. Beyond coding theory, related problems of noisy polynomial interpolation and polynomial reconstruction have appeared in cryptography, both as hardness assumptions and as building blocks for cryptographic protocols. Examples include oblivious polynomial evaluation~\cite{naor1999} and noisy polynomial interpolation over finite fields~\cite{bleichenbacher2000}. 

\section{Background}

\subsection{Notation}
We summarize the key notation used throughout this paper:
\begin{center}
\begin{tabular}{ll}
\toprule
\textbf{Symbol} & \textbf{Description} \\
\midrule
$\GF{q}$, $\GF{q^2}$ & Finite field of size $q$, extension field of size $q^2$ \\
$n$ & Number of constraints / code length ($n = q^3$ for Hermitian) \\
$k$ & Number of variables / code dimension \\
$d$, $d^\perp$ & Minimum distance of code and dual code \\
$\ell$ & Maximum errors efficiently decodable \\
\midrule
$\mathcal{H}$ & Hermitian curve: $y^q + y = x^{q+1}$ \\
$\mathcal{P}$ & Rational points on $\mathcal{H}$: $\{P_1, \ldots, P_n\}$ \\
$\mathcal{L}_t$ & Rational functions with pole degree $\leq t$ at infinity \\
$\mathcal{C}_t$ & Hermitian code with parameter $t$ \\
$g$ & Genus of Hermitian curve: $q(q-1)/2$ \\
\midrule
$F_i$ & Constraint set at position $i$ \\
$r$ & Size of constraint sets \\
$\expval{s}$ & Expected number of satisfied constraints \\
\bottomrule
\end{tabular}
\end{center}

\subsection{Decoded Quantum Interferometry}

Decoded Quantum Interferometry (DQI) is a quantum algorithm for combinatorial optimization that exploits a connection between optimization problems and error-correcting codes. The algorithm targets a broad class of constraint satisfaction problems called max-LINSAT, which we now define.

Given a prime $p$, integers $n > k$, and a matrix $B \in \GF{p}^{n \times k}$, consider $n$ constraints of the form $\mathbf{b}_i \cdot \mathbf{x} \in F_i$, where $\mathbf{b}_i$ is the $i$-th row of $B$ and $F_i \subset \GF{p}$ is an arbitrary subset. The max-LINSAT problem seeks $\mathbf{x} \in \GF{p}^k$ satisfying as many constraints as possible. This formulation captures many important optimization problems, including polynomial regression over finite fields and various constraint satisfaction problems. To simplify notation, we associate with each constraint a function $f_i: \GF{p} \to \{+1, -1\}$ where $f_i(y) = +1$ if $y \in F_i$ and $f_i(y) = -1$ otherwise. The optimization objective becomes
\begin{equation}
f(\mathbf{x}) = \sum_{i=1}^n f_i(\mathbf{b}_i \cdot \mathbf{x})
\end{equation}
which counts satisfied constraints minus unsatisfied constraints.

The central idea of DQI is to prepare quantum states that are biased toward solutions with high objective values. Specifically, the algorithm creates states of the form
\begin{equation}
\ket{P(f)} \propto \sum_{\mathbf{x} \in \GF{p}^k} P(f(\mathbf{x})) \ket{\mathbf{x}}
\end{equation}
where $P$ is a degree-$\ell$ polynomial chosen to amplify states corresponding to large $f(\mathbf{x})$ values. Measuring this state in the computational basis yields samples biased toward more optimal solutions.

As shown in \cite{dqi}, these states can be prepared efficiently through a reduction to syndrome decoding. The algorithm interprets the matrix $B$ as defining a linear code $\mathcal{C} = \{ B \mathbf{x} : \mathbf{x} \in \GF{p}^k\}$, specifically working with the dual code $\mathcal{C}^\perp = \{\mathbf{d} \in \GF{p}^n : B^T \mathbf{d} = \mathbf{0}\}$. The preparation of $\ket{P(f)}$ requires solving syndrome decoding problems for $\mathcal{C}^\perp$ with up to $\ell \leq \lfloor \frac{d^\perp - 1}{2} \rfloor$ errors (where $d^\perp$ is the distance of the dual code), implemented reversibly in quantum superposition.

The DQI algorithm proceeds through five main steps. First, it prepares superpositions over Dicke states (uniform superpositions of computational basis states with fixed Hamming weight). Second, it applies phases determined by the constraint vector to encode problem-specific information. Third, it computes syndromes through reversible matrix-vector multiplication. Fourth, it performs the crucial uncomputation step by solving syndrome decoding problems to remove entanglement between error patterns and syndromes. Finally, it applies a quantum Fourier transform to obtain the desired optimization state.

The algorithm's performance is governed by the semicircle law, which provides a formula for the expected number of satisfied constraints in the asymptotic limit. For instance, in the balanced case where $|F_i| = p/2$ for all constraints (meaning each constraint is satisfied by exactly half the field elements), the expected number of constraints $\expval{s}$ that DQI satisfies is
\begin{equation}
\frac{\expval{s}}{n} = \frac{1}{2} + \sqrt{\frac{\ell}{n}\left(1 - \frac{\ell}{n}\right)},
\end{equation}
where $\ell$ is the maximum number of errors that can be efficiently decoded in $\mathcal{C}^\perp$ ($\ell$ is upper bounded by $\lfloor \frac{d-1}{2}\rfloor$). This formula reveals that DQI's optimization performance is fundamentally limited by the error-correcting capabilities of the underlying code family, establishing a direct bridge between coding theory and quantum optimization.

\subsection{Hermitian Codes}\label{sec:herm-codes}
Hermitian codes were introduced to address a fundamental limitation in classical coding theory: the trade-off between code length and performance. Reed-Solomon codes, while optimal in many respects, are limited in length to at most the field size. For a field $\GF{q}$, Reed-Solomon codes can have length at most $q$ because they evaluate univariate polynomials at distinct field elements, and there are only $q$ such elements available. For applications requiring longer codes with good error-correcting properties, this constraint becomes restrictive.

Beyond their excellent error-correcting parameters, Hermitian codes offer a crucial computational advantage over Reed-Solomon codes for quantum implementations. For a given block length $n$, Reed-Solomon codes require field size at least $n$, necessitating $\lceil\log_2(n)\rceil$ qubits to represent each field element. In contrast, Hermitian codes achieve block length $n = q^3$ using the smaller field $\GF{q^2}$, requiring only $\lceil 2\log_2(q)\rceil = \lceil(2/3)\log_2(n)\rceil$ qubits per field element. This approximate one-third reduction in qubit requirements per field element makes Hermitian codes particularly attractive for near-term quantum implementations of DQI.

Hermitian codes overcome this limitation through a clever geometric insight: instead of evaluating univariate polynomials at points on a line (which gives at most $q$ points), they evaluate bivariate rational functions at points on an algebraic curve in the plane. The Hermitian curve over $\GF{q^2}$ contains $q^3$ points, yielding codes of length $q^3$ using a field of size $q^2$---a substantial improvement over the Reed-Solomon bound.

Hermitian codes offer several compelling advantages for DQI applications. First, their use of smaller alphabet sizes compared to Reed--Solomon codes of equivalent length translates to more efficient quantum implementations, since representing large field elements with qubits is generally expensive. This may make HOPI an attractive candidate for verifiable quantum advantage on fault-tolerant quantum computers of moderate size.

Second, studying DQI with Hermitian codes provides insight into the broader applicability of the framework, helping identify which structural properties of algebraic codes are essential for quantum optimization advantages.

Third, HOPI can be viewed as a problem of approximate list recovery for Hermitian codes. Recall that both the dual code $\mathcal{C}^\perp$ and primal code $\mathcal{C}$ are Hermitian. Because $\mathcal{C}$ is Hermitian the HOPI problem can be cast as finding a codeword in a Hermitian code that obeys as many list constraints as possible. This is a well-known problem called list-recovery (see e.g. \cite{GSI23}). This problem can arise, for example, if large alphabet codes are transmitted in binary, in which case losing one bit from the string representing a symbol yields a list size of half the alphabet. Maximizing agreement with large sets is equivalent to minimizing Hamming distance to an affine code coset, a problem known to be computationally intractable in general and hard to approximate~\cite{BerlekampMcElieceTilborg1978,dumer2003}. 

Hermitian codes are constructed by evaluating rational functions at points on the Hermitian curve over finite fields. The Hermitian curve over $\GF{q^2}$ is defined by the equation
\begin{equation}\label{eq:hermite}
y^q + y = x^{q+1}
\end{equation}
where $q$ is a prime power. This curve determines a specific set of points in the plane $\GF{q^2} \times \GF{q^2}$ that will serve as evaluation points for the code construction. The Hermitian curve is remarkable because it has exactly $q^3$ solutions, which is the maximum possible for any smooth curve of its genus over $\GF{q^2}$~\cite{stichtenoth1993}.

To construct Hermitian codes, we evaluate rational functions at these curve points. A rational function is a quotient $f(x,y) = p(x,y)/r(x,y)$ where $p$ and $r$ are polynomials. However, since we want the evaluation to be well-defined at our chosen points, we restrict to functions that are finite and well-defined at all evaluation points. More concretely, let $\mathcal{P} = \{P_1, P_2, \ldots, P_n\}$ be the set of $n = q^3$ rational points on the Hermitian curve. For a parameter $t$, we consider the vector space $\mathcal{L}_t$ of rational functions that have poles of total degree at most $t$ (where poles may occur only away from the evaluation set $\mathcal{P}$) and are regular (finite and well-defined) at all points in $\mathcal{P}$ (see \cref{app:rr}). The Hermitian code $\mathcal{C}_t$ consists of all codewords obtained by evaluating functions from $\mathcal{L}_t$:
\begin{equation}
\mathcal{C}_t = \{(f(P_1), f(P_2), \ldots, f(P_n)) : f \in \mathcal{L}_t\}
\end{equation}
Due to the restriction that all functions in $\mathcal{L}_t$ must be well-defined at all points in $\mathcal{P}$, it essentially consists of bivariate polynomials in the affine coordinates $x$ and $y$ of the Hermitian curve.

The parameters of these codes are determined by fundamental results from algebraic geometry. The code length is $n = q^3$, fixed by the number of evaluation points on the curve. The dimension grows as $k = t + 1 - g$ where $g = q(q-1)/2$ is the genus of the curve. The genus $g$ is a fundamental invariant measuring the topological complexity of the curve; intuitively, it counts the number of ``holes'' or ``handles'' the curve would have as a surface. The minimum distance satisfies $d \geq n - t$. These parameters yield codes with exceptional performance. For appropriate choices of $t$, Hermitian codes achieve nearly optimal trade-offs between information rate $k/n$ and error-correcting capability $d/n$, often approaching or meeting the Singleton bound that limits all linear codes. In many parameter regimes, they outperform Reed-Solomon codes of comparable length.

\paragraph{Dual Codes.} Remarkably, the dual of a Hermitian code is again a Hermitian code on the same curve, but with a different pole order parameter~\cite{stichtenoth1993}. More precisely, if $\mathcal{C}_t$ denotes the Hermitian code obtained from functions with pole order at most $t$ at the point at infinity, then
\begin{equation}
\mathcal{C}_t^\perp = \mathcal{C}_{t'} \qquad \text{with} \quad t' = n + 2g - 2 - t,
\end{equation}
where $n = q^3$ is the code length and $g = q(q-1)/2$ is the genus of the Hermitian curve. This duality relation arises from the Riemann–Roch theorem and the canonical divisor of the curve (see \cref{app:rr}). In practice, this means that efficient decoding algorithms exist not only for the original code $\mathcal{C}_t$, \emph{but also for its dual}. For DQI this feature is crucial: the syndrome decoding step always involves the dual code, and Hermitian codes provide the guarantee that dual decoding is no harder than primal decoding.

\paragraph{Decoding Algorithms.} Decoding algorithms for Hermitian codes have been developed along several complementary lines, and provide the crucial foundation that makes DQI applicable in this setting. The earliest methods were direct generalizations of the classical syndrome-based approach used for Reed–Solomon codes. In this paradigm one computes a syndrome vector, then solves a so-called ``key equation'' that relates the error locator and evaluator polynomials. For Hermitian codes this step is performed using Sakata's multidimensional generalization of the Berlekamp–Massey algorithm~\cite{sakata1990,sakata1995}, which guarantees unique decoding up to $\lfloor (d-1)/2 \rfloor$ errors in time polynomial in the block length.

A different but related family of algorithms is based on the Feng–Rao, or majority-coset, idea~\cite{fengrao}. These methods exploit the large automorphism group of the Hermitian curve to recover additional syndromes and thereby extend the guaranteed decoding radius, in some cases beyond the designed minimum distance. More recent refinements of this approach have clarified how it applies to both the code and its dual~\cite{geil2013}.

Beyond unique decoding, Hermitian codes also admit list-decoding algorithms that can in principle correct far beyond half the minimum distance. The most widely used of these are interpolation-based methods in the spirit of Guruswami–Sudan~\cite{gs1998}, adapted to the Hermitian setting by Lee and O'Sullivan through a Gröbner-basis formulation~\cite{lee2006}. Later work extended these ideas to broader classes of algebraic geometry codes and improved their efficiency~\cite{matsumoto2013,nielsen2015}. Extending even further, \emph{soft-decision} decoding algorithms can exploit reliability information about received symbols to achieve better performance than hard-decision methods. The Koetter-Vardy algorithm~\cite{koetter2003} extends Guruswami--Sudan list decoding to incorporate soft information by using symbol reliabilities to determine interpolation multiplicities, achieving significant coding gains for Reed--Solomon codes. This approach has been successfully extended to algebraic geometry codes~\cite{chen2009}, including Hermitian codes, where the soft information is incorporated into the polynomial interpolation step over the function field.

For our purposes, the most important point is not the detailed taxonomy of algorithms, but the fact that Hermitian codes and their duals both admit efficient classical decoding with well-understood guarantees. This ensures that the syndrome-decoding step required by DQI can be implemented in principle by reversible circuits, providing the bridge from algebraic geometry to quantum optimization.

\subsection{Hermitian Optimal Polynomial Intersection}
The Hermitian Optimal Polynomial Intersection (HOPI) problem is a direct generalization of the Reed-Solomon case, inheriting the constraint structure that enables quantum advantage while exploiting the superior parameters of Hermitian codes.

Given the Hermitian curve $\mathcal{H}: y^q + y = x^{q+1}$ over $\GF{q^2}$, let $\mathcal{P} = \{P_1, P_2, \ldots, P_n\}$ be the set of rational points on $\mathcal{H}$ excluding the point at infinity, so $n = q^3$. For each point $P_i \in \mathcal{P}$, we are given a subset $F_i \subset \GF{q^2}$ of allowed values. The HOPI problem seeks a rational function $f \in \mathcal{L}_t$ (the space of rational functions with poles of degree at most $t$ at infinity) that maximizes the objective
\begin{equation}
\text{maximize} \quad |\{i : f(P_i) \in F_i\}|
\end{equation}
An important parameter of the optimization problem is the set size $r = \max_i \abs{F_i}$. In this work, as in Ref.~\cite{dqi}, we will work in the `large set' regime, wherein $r$ is proportional to the field size.

This directly generalizes the Reed-Solomon OPI problem from univariate polynomials evaluated at field elements to bivariate rational functions evaluated at points on an algebraic curve. Where Reed-Solomon OPI seeks polynomials $p(x)$ of degree less than $n$ that maximize agreement with constraint sets at field elements, HOPI seeks rational functions on the Hermitian curve that maximize agreement with constraint sets at curve points. The geometric richness of the Hermitian curve enables $q^3$ constraints over fields of size $q^2$, compared to Reed-Solomon's limitation to at most $q$ constraints over fields of size $q$.

The HOPI problem admits a natural interpretation as a list-recovery problem for Hermitian codes, connecting it directly to classical algebraic decoding algorithms. In coding theory, list-recovery seeks all codewords that simultaneously belong to given constraint sets at specified positions. For HOPI, each rational function $f \in \mathcal{L}_t$ corresponds to the codeword $(f(P_1), f(P_2), \ldots, f(P_n))$ in the associated Hermitian code, making HOPI equivalent to finding the codeword that agrees with the largest number of constraint sets.

This list-recovery perspective illuminates why HOPI is computationally challenging for classical algorithms. Interpolation–based list-decoding and list-recovery methods (in the spirit of Guruswami–Sudan) are effective when each input list is small relative to the alphabet and when the sought codeword agrees with a large fraction of positions. In our setting the lists are \emph{large}: $r/|\GF{q^2}|$ is a constant bounded away from zero (balanced cases even have $r\approx|\GF{q^2}|/2$), so the algorithm must tolerate list sizes $\Theta(q^2)$ per coordinate. Known AG list-recovery algorithms either require $r$ to be sub-alphabetic (e.g., polylogarithmic in the alphabet size) or need agreement well above the random baseline to guarantee a small output list; in the large-set regime the guaranteed agreement thresholds are not met and, in practice, one faces either an empty output or a combinatorial blow-up in the number of candidates. Moreover, pushing interpolation multiplicities high enough to handle large lists causes the underlying Gröbner-basis/module-minimization steps to become prohibitively expensive. Thus, the classical list-recovery toolbox does not yield efficient algorithms with nontrivial guarantees in precisely the large-set regime where DQI's bias (mediated by dual decoding) remains effective.

Moreover, the list-recovery interpretation reveals why HOPI inherits the favorable properties needed for DQI. The dual of a Hermitian code is itself a Hermitian code with efficient decoding algorithms, providing exactly the syndrome decoding capability that DQI requires. The problem structure ensures that classical optimization faces the same fundamental difficulty as classical list-recovery in the large-constraint regime, while quantum interference can exploit the underlying algebraic structure to achieve improved performance.

\section{Results}

The Hermitian Optimal Polynomial Intersection problem is an instance of the general max-LINSAT problem over the extension field $\GF{q^2}$. The performance of DQI on such problems is governed by the semicircle law established in~\cite{dqi}. That analysis requires only the linearity of the underlying dual code and the existence of an efficient decoder up to some error radius~$\ell$, and therefore applies verbatim once the code family is fixed. Since the dual of a Hermitian code is again a Hermitian code with well-defined minimum distance, and since efficient decoders exist at least up to half that distance (see~\cref{sec:herm-codes}), the framework carries over directly.

In the Hermitian construction with parameter $t$, the block length is $n=q^3$, the dimension is $k=\dim \mathcal{L}_t$, and the minimum distance satisfies $d \geq n-t$. The dual is again a Hermitian code with parameter $t' = n+2g-2-t$, where $g=q(q-1)/2$ is the genus, and thus the dual distance $d^\perp$ provides the relevant decoding radius. Throughout we assume $2\ell+1<d^\perp$, so that decoding up to $\ell=\lfloor (d^\perp-1)/2 \rfloor$ errors is guaranteed, and we use this value of~$\ell$ in applying the semicircle law.

\begin{theorem}[DQI performance for HOPI]\label{thm:hopi-semicircle}
Let $q$ be a prime power and consider an instance of HOPI over the Hermitian curve with
$n=q^3$ constraints, alphabet $\GF{q^2}$, and constraint sets of size $r$. 
Let $\mathcal{C}^\perp$ be the dual Hermitian code of minimum distance $d^\perp$, and assume efficient decoding up to 
$\ell = \lfloor (d^\perp-1)/2 \rfloor$ errors (recall that for a Hermitian code $\mathcal{C}_t$, $d^\perp \geq t+2-2g$). Then Decoded Quantum Interferometry produces solutions with expected satisfaction fraction
\begin{equation}
\frac{\expval{s}_{\mathrm{DQI}}}{n}
=
\begin{cases}
\displaystyle
\left(\sqrt{\tfrac{\ell}{n}\!\left(1-\tfrac{r}{q^2}\right)}+
\sqrt{\tfrac{r}{q^2}\!\left(1-\tfrac{\ell}{n}\right)}\right)^{\!2},
& \text{if } \tfrac{r}{q^2}\le 1-\tfrac{\ell}{n},\\[1.2ex]
1, & \text{otherwise}.
\end{cases}
\end{equation}
\end{theorem}
This statement makes precise the connection between the error-correcting capability of the dual Hermitian code and the quality of approximate solutions obtainable by DQI. The only nontrivial inputs are the length $n=q^3$, the constraint-set size $r$, and the decoding radius $\ell$, which in turn comes from the dual distance. With these parameters in hand, the performance follows directly from the general semicircle law.

To quantify the advantage, we compare to Prange’s information set decoding~\cite{prange}, following the same baseline choice as in~\cite{dqi}. In this context we take $k=\dim \mathcal{L}_t$. Prange’s algorithm guarantees satisfaction of $k$ constraints exactly, and in expectation satisfies a fraction $r/q^2$ of the remaining $n-k$. Its expected performance is therefore
\begin{equation}
    \frac{\expval{s}_{\mathrm{Prange}}}{n} = \frac{k + (r/q^2)\cdot(n-k)}{n}.
\end{equation}
Polynomially many repetitions improve this fraction only by $o(1)$, so this expression accurately captures its asymptotic behavior. Local search heuristics such as simulated annealing perform only marginally better than random guessing in the dense, highly structured setting of HOPI and are not competitive.

\begin{figure}[H]
    \centering
    \includegraphics[width=0.75\linewidth]{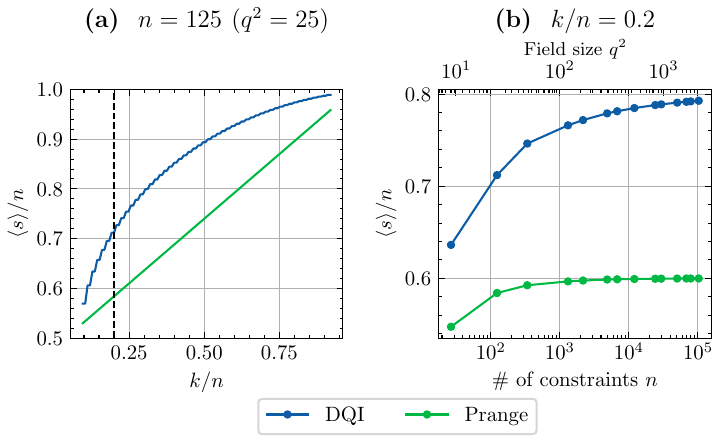}
    \caption{Performance of DQI on HOPI. Panel (a) shows the balanced case ($r=q^2/2$) for $q^2=25$ ($n=125$), plotting the expected satisfaction fraction $\expval{s}/n$ against code rate $k/n$. DQI follows the semicircle curve, while Prange's performance increases linearly. Panel (b) fixes $k/n=0.2$ and shows performance as $q$ grows, confirming that the quantum advantage persists at large $n$.}
    \label{fig:adv1}
\end{figure}

\Cref{fig:adv1} shows the resulting comparison for the balanced case $r=q^2/2$. Panel (a) plots the expected satisfaction fraction as a function of rate $k/n$ for a fixed field size $q^2=25$ ($n=125$). DQI follows the semicircle curve and consistently outperforms Prange’s linear profile across all rates. Panel (b) fixes a representative rate $k/n=0.2$ and increases the field size. The advantage persists and even grows with $n$, indicating that the separation is not a small-size artifact but a scalable feature of the framework.

Beyond this baseline, we also implemented simulated annealing and found it to perform strictly worse than Prange across the parameter ranges we tested (rates $k/n \in [0.1,0.5]$, with balanced $r$). This is consistent with intuition for dense, highly constrained instances: single- or few-coordinate moves change the objective by $O(1)$ clauses out of $n=\Theta(q^3)$, so the energy landscape seen by local search is nearly flat around the random baseline. Since Prange hard-codes exact satisfaction of $k$ constraints before random completion, its expected fraction dominates local-search heuristics that cannot reliably enforce those $k$ equalities.

\begin{figure}[H]
    \centering
    \includegraphics[width=0.5\linewidth]{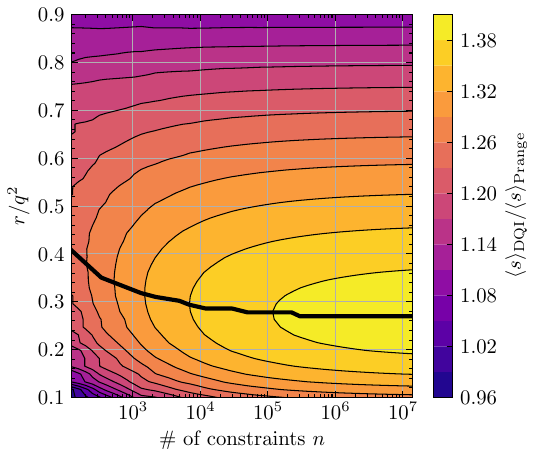}
    \caption{Advantage ratio $\expval{s}_{\mathrm{DQI}}/\expval{s}_{\mathrm{Prange}}$ as a function of $n$ and $r$. The rate is fixed at $k/n=0.2$. The black line marks the constraint-set size $r$ at which the ratio is maximized, occurring around $r/q^2\approx 0.28$.}
    \label{fig:adv2}
\end{figure}

To explore the broader parameter space, \Cref{fig:adv2} plots the advantage ratio $\expval{s}_{\mathrm{DQI}}/\expval{s}_{\mathrm{Prange}}$ as a function of both $n$ and $r$. Two trends emerge clearly. First, the advantage over Prange grows as $n$ increases. Second, the ratio is maximized not at the balanced point but for unbalanced constraint sets around $r/q^2 \approx 0.28$. In this regime DQI satisfies nearly 40\% more constraints compared to Prange (in expectation). These curves are computed directly from the closed-form expressions above; no heuristic assumptions beyond those stated enter into the analysis.

In summary, once the Hermitian code parameters are substituted into the general semicircle law, DQI yields a satisfaction fraction that scales favorably with problem size and consistently dominates Prange’s baseline. The performance advantage is quantitatively linked to the dual distance of the Hermitian code, reinforcing the central message that DQI’s power derives from efficient decoding of structured codes rather than any property unique to Reed–Solomon.

\section{Limits and Open Questions}

Our extension of DQI to Hermitian codes provides evidence that the quantum advantage observed in the original Reed--Solomon work is not specific to a single code family, but rather a more general feature of the framework. By demonstrating that the advantage arises from the ability to exploit algebraic structure rather than from idiosyncrasies of Reed--Solomon codes, this work strengthens the case for DQI as a general principle for achieving quantum advantage in structured optimization problems. However, several important limitations and open questions remain.

The first open question regards whether the approximate optima efficiently achieved by DQI on the HOPI problem can be improved even further through the use of list decoding algorithms for decoding Hermitian codes beyond $\lfloor (d^\perp-1)/2 \rfloor$ errors. While list-decoding methods such as Guruswami--Sudan extend in principle to Hermitian codes, their efficient implementation in the parameter regime relevant to HOPI are less well understood. Additionally, when decoding beyond $\lfloor (d^\perp-1)/2 \rfloor$ one must bound the effects of decoding failure on DQI, as discussed in~\cite{dqi}. It remains open whether there exist decoders for Hermitian codes that approach the information-theoretic decoding radius, and more quantitatively, whether such algorithms can be made reversible with low overhead for coherent use in DQI on quantum computers with reasonable resources.

Another open question is how far the framework extends beyond Hermitian codes. Although Hermitian codes are among the most well-studied algebraic geometry codes and provide an especially clean duality structure, they are only one family. A related question concerns whether the field size reduction achieved by Hermitian codes --- approximately one-third fewer qubits per field element compared to Reed-Solomon codes --- can be improved further with other code families. Binary codes such as Goppa codes operate over $\GF{2}$~\cite{Goppa1970}, providing maximal field representation efficiency, but it remains unclear whether such codes possess the necessary duality properties and efficient dual decoding algorithms required for DQI. More generally, exploring the trade-off between field size reduction and the algebraic structure needed for effective quantum interference represents an important direction for optimizing the practical implementation of DQI on near-term quantum devices.

Finally, as with all apparent quantum speedups for non-oracular problems, the possibility remains that a more powerful classical algorithm will be discovered that outperforms it. We pose this as a challenge for the classical algorithms community.

In summary, while the extension of DQI to Hermitian codes demonstrates that the framework is not limited to Reed--Solomon codes and confirms that quantum advantage can be achieved through exploiting algebraic structure, it also leaves open significant questions about decoding efficiency, generalization to other algebraic geometry codes, and the development of even more powerful classical or quantum methods. These questions define important directions for future work in quantum optimization and algebraic coding theory. \\

\noindent
\textbf{Acknowledgments:} We thank Robbie King, Greg Kuperberg, and Mary Wootters for helpful discussions, and we thank Noah Shutty for his comments on earlier drafts of this work.

\bibliography{refs}

\clearpage

\appendix

\section{Basics of Algebraic Geometry over Finite Fields}

This appendix provides the algebraic geometry background needed to understand the construction and properties of Hermitian codes. We work throughout over finite fields, though most definitions extend to arbitrary fields. The presentation follows standard references, particularly Stichtenoth~\cite{stichtenoth1993}. The connection between algebraic geometry and error-correcting codes arises from a simple but powerful idea: encode information by evaluating polynomials or rational functions at specified points, then use the algebraic structure of these functions to design efficient decoding algorithms. Classical Reed-Solomon codes implement this strategy using univariate polynomials evaluated at points in a finite field, while algebraic geometry codes generalize this approach using rational functions on algebraic curves.

\subsection{Algebraic Varieties and Curves}

\subsubsection{Affine and Projective Space}

We begin with the most basic geometric objects: affine and projective spaces over finite fields.

\begin{definition}[Affine space]
The affine $n$-space over a field $\Bbbk$, denoted $\mathbb{A}^n(\Bbbk)$, is the set of all $n$-tuples of elements in $\Bbbk$:
\begin{equation}
    \mathbb{A}^n(\Bbbk) = \{(a_1, \ldots, a_n) \mid a_i \in \Bbbk\}.
\end{equation}
\end{definition}

Affine space provides a natural setting for polynomial equations, but it lacks certain desirable properties. Most importantly for our applications, it has no natural notion of ``points at infinity,'' which creates technical difficulties when studying rational functions and their poles.

\begin{definition}[Projective space]
The projective $n$-space over $\Bbbk$, denoted $\mathbb{P}^n(\Bbbk)$, is the set of lines through the origin in $\mathbb{A}^{n+1}(\Bbbk)$. Equivalently, it consists of all tuples $(a_0,\ldots,a_n) \in \mathbb{A}^{n+1}(\Bbbk) \setminus \{(0,\ldots,0)\}$ under the equivalence relation $(a_0,\ldots,a_n) \sim (\lambda a_0, \ldots, \lambda a_n)$ for any $\lambda \neq 0$. We write points in $\mathbb{P}^n(\Bbbk)$ using homogeneous coordinates $[a_0 : \cdots : a_n]$.
\end{definition}

Projective space ``compactifies'' affine space by adding points at infinity. The affine $n$-space embeds naturally into $\mathbb{P}^n(\Bbbk)$ via $(a_1, \ldots, a_n) \mapsto [1 : a_1 : \cdots : a_n]$. Points with $a_0 = 0$ correspond to directions ``at infinity'' in the affine space.

\subsubsection{Algebraic Varieties}

\begin{definition}[Affine variety]
Let $S$ be a set of polynomials in $\Bbbk[x_1,\ldots,x_n]$. The affine variety defined by $S$, denoted $V(S)$, is the set of common zeros of all polynomials in $S$:
\begin{equation}
    V(S) = \{P \in \mathbb{A}^n(\Bbbk) \mid f(P) = 0 \text{ for all } f \in S\}.
\end{equation}
\end{definition}

For projective varieties, we must restrict to homogeneous polynomials to ensure that the zero condition is well-defined on equivalence classes. A polynomial $f(x_0, \ldots, x_n)$ is homogeneous of degree $d$ if every monomial has total degree $d$. If $f$ is homogeneous of degree $d$ and $f(a_0, \ldots, a_n) = 0$, then $f(\lambda a_0, \ldots, \lambda a_n) = \lambda^d f(a_0, \ldots, a_n) = 0$ for any $\lambda \neq 0$.

\begin{definition}[Irreducible variety]
A variety is irreducible if it cannot be written as the union of two proper subvarieties.
\end{definition}

An algebraic curve is an irreducible variety of dimension 1. For coding applications, we focus on smooth projective curves over finite fields.

\begin{definition}[Smooth curve]
A curve is smooth (or nonsingular) if it has no singular points. A point $P$ on a curve defined by $f(x,y) = 0$ is singular if both partial derivatives $\partial f/\partial x$ and $\partial f/\partial y$ vanish at $P$.
\end{definition}

\subsubsection{Function Fields and Rational Functions}

Given an irreducible affine variety $V \subseteq \mathbb{A}^n$, we can study polynomial and rational functions on $V$.

\begin{definition}[Function field]
Let $V$ be an irreducible variety and let $I(V)$ denote the ideal of polynomials vanishing on $V$. The coordinate ring is $\Bbbk[V] = \Bbbk[x_1,\ldots,x_n]/I(V)$. The function field of $V$, denoted $\Bbbk(V)$, is the field of fractions:
\begin{equation}
    \Bbbk(V) = \left\{\frac{f}{g} \mid f, g \in k[V], g \neq 0\right\}.
\end{equation}
\end{definition}

Elements of $\Bbbk(V)$ are rational functions on $V$. The condition $g \neq 0$ means $g$ is not identically zero on $V$; it may still have zeros at individual points.

\subsection{Divisors and Riemann-Roch Theory}

Divisors provide a systematic way to encode information about zeros and poles of rational functions.

\subsubsection{Divisors}

\begin{definition}[Divisor]
A divisor on a curve $\mathcal{X}$ is a formal finite sum
\begin{equation}
    D = \sum_{P \in \mathcal{X}} n_P P,
\end{equation}
where $n_P \in \mathbb{Z}$ and only finitely many $n_P$ are nonzero. The degree of $D$ is $\deg(D) = \sum_P n_P$. We write $D \geq 0$ if all coefficients $n_P \geq 0$.
\end{definition}
The support of a divisor $D = \sum_P n_P P$, denoted $\text{supp}(D)$, is the set of points $P$ where $n_P \neq 0$.

\begin{definition}[Principal divisor]
For a nonzero rational function $f \in k(\mathcal{X})$, the principal divisor $(f)$ is
\begin{equation}
    (f) = \sum_{P \in \mathcal{X}} \text{ord}_P(f) \cdot P,
\end{equation}
where $\text{ord}_P(f)$ is the order of $f$ at $P$: positive if $f$ has a zero at $P$, negative if $f$ has a pole at $P$, and zero if $f$ is nonzero and finite at $P$. For example, if $f(x) = (x-a)^2/(x-b)^3$, then $\text{ord}_a(f) = 2$ (a double zero) and $\text{ord}_b(f) = -3$ (a triple pole), then $(f) = 2a-3b$.
\end{definition}

A fundamental property of curves is that rational functions have the same number of zeros and poles, counted with multiplicity.

\begin{theorem}[Degree zero property]
For any nonzero rational function $f$ on a smooth projective curve $\mathcal{X}$, $\deg((f)) = 0$.
\end{theorem}

This generalizes the familiar fact from complex analysis that a meromorphic function on the Riemann sphere has equally many zeros and poles.

\subsubsection{Riemann-Roch Spaces}\label{app:rr}

The Riemann-Roch spaces encode rational functions with prescribed pole behavior.

\begin{definition}[Riemann-Roch space]
For a divisor $D$ on a smooth projective curve $\mathcal{X}$, the Riemann-Roch space is
\begin{equation}
    L(D) = \{f \in \Bbbk(\mathcal{X}) \mid (f) + D \geq 0\} \cup \{0\}.
\end{equation}
\end{definition}

Intuitively, $L(D)$ consists of rational functions whose poles are ``no worse'' than those allowed by $D$. If $D = \sum_P n_P P$ with $n_P > 0$, then functions in $L(D)$ may have poles of order at most $n_P$ at point $P$. The space $L(D)$ is a finite-dimensional vector space over $\Bbbk$, with dimension denoted $\ell(D)$.

\begin{theorem}[Riemann-Roch theorem]
Let $\mathcal{X}$ be a smooth projective curve of genus $g$, and let $K$ be a canonical divisor of degree $2g-2$ (whose precise definition involves differential forms, but for our purposes only the degree matters). Then for any divisor $D$,
\begin{equation}
    \ell(D) - \ell(K - D) = \deg(D) - g + 1.
\end{equation}
\end{theorem}

\begin{corollary}
If $\deg(D) > 2g - 2$, then $\ell(K - D) = 0$ and
\begin{equation}
    \ell(D) = \deg(D) - g + 1.
\end{equation}
\end{corollary}

The genus $g$ is a fundamental invariant measuring the ``complexity'' of the curve. For example, $g = 0$ for rational curves (isomorphic to the projective line), and $g = 1$ for elliptic curves.

\subsection{Algebraic Geometry Codes}\label{app:ag-code}

We now construct linear codes by evaluating rational functions at points on algebraic curves. 

\subsubsection{General Construction}

\begin{definition}[Algebraic geometry code]
Let $\mathcal{X}$ be a smooth projective curve over $\GF{q}$, let $\mathcal{P} = \{P_1, \ldots, P_n\}$ be $n$ distinct rational points on $\mathcal{X}$, and let $G$ be a divisor with support disjoint from $\mathcal{P}$. The evaluation map is
\begin{equation}
    \text{ev}_{\mathcal{P}}: L(G) \to \GF{q}^n, \quad f \mapsto (f(P_1), \ldots, f(P_n)).
\end{equation}
The algebraic geometry code $C_L(\mathcal{X}, \mathcal{P}, G)$ is the image of this map. We write $\mathcal L_t := L(tP_\infty)$ and $\mathcal C_t := C_L(\mathcal H,\mathcal P,tP_\infty)$ to match the main text.
\end{definition}
The requirement $\text{supp}(G) \cap \mathcal{P} = \emptyset$ ensures that functions in $L(G)$ are finite and well-defined at all evaluation points: we cannot evaluate a function at a point where it has a pole.

The parameters of $C_L(\mathcal{X}, \mathcal{P}, G)$ are determined by the geometry:
\begin{itemize}
    \item \textbf{Length}: $n = |\mathcal{P}|$ (number of evaluation points)
    \item \textbf{Dimension}: $k = \ell(G)$ when $\deg(G) < n$ (ensuring the evaluation map is injective, so distinct functions yield distinct codewords)
    \item \textbf{Minimum distance}: $d \geq n - \deg(G)$ (Singleton-type bound)
\end{itemize}
When $\deg(G) > 2g - 2$, the Riemann-Roch theorem gives $k = \deg(G) - g + 1$. Reed-Solomon codes provide the simplest example of this construction: take $\mathcal{X} = \mathbb{P}^1$ (the projective line), $\mathcal{P} = \{a_1, \ldots, a_n\} \subset \GF{q}$, and $G = m \cdot P_\infty$ where $P_\infty$ is the point at infinity. Then $L(G)$ consists of polynomials of degree at most $m$, and evaluation at $\mathcal{P}$ gives the classical Reed-Solomon construction.

\subsubsection{Duality for Algebraic Geometry Codes}

A remarkable feature of algebraic geometry codes is that their duals have a clean geometric description.

\begin{theorem}[Duality for AG codes]
The dual of the algebraic geometry code $C_L(\mathcal{X}, \mathcal{P}, G)$ is isomorphic to another algebraic geometry code of the form $C_L(\mathcal{X}, \mathcal{P}, G')$ for an explicitly computable divisor $G'$.
\end{theorem}

The specific form of $G'$ depends on technical details involving canonical divisors and differentials, but the key point is that \emph{both the code and its dual admit the same type of algebraic structure}. This ensures that efficient decoding algorithms designed for one can often be adapted to the other. This duality structure is crucial for practical applications: many efficient decoding algorithms work by exploiting the algebraic relationships between a code and its dual, so having both codes in the same well-understood family greatly simplifies algorithm design.

\subsection{Hermitian Codes}\label{app:hermitian}

Hermitian codes, constructed from the Hermitian curve, provide some of the best-known examples of algebraic geometry codes.

\subsubsection{The Hermitian Curve}

\begin{definition}[Hermitian curve]
The Hermitian curve over $\GF{q^2}$ is the projective curve defined by
\begin{equation}
    y^q z + y z^q = x^{q+1}.
\end{equation}
In affine coordinates (setting $z = 1$), this becomes $y^q + y = x^{q+1}$.
\end{definition}

The Hermitian curve has several exceptional properties. First, it contains many rational points: exactly $q^3$ affine rational points over $\GF{q^2}$, plus one point at infinity (the point at infinity is $P_\infty = [0:1:0]$ in projective coordinates). Second, it has high genus $g = q(q-1)/2$, which grows quadratically with the field parameter. Finally, the curve is maximal in the sense that its number of rational points achieves the Hasse-Weil upper bound for curves of this genus. These properties make Hermitian curves ideal for coding: the large number of rational points ($q^3$) provides long codes, while the relatively small field ($\GF{q^2}$ rather than $\GF{q^3}$) keeps the alphabet size manageable for efficient implementation.

\subsubsection{Construction and Parameters}

Hermitian codes are constructed by taking $G = m P_\infty$ for various integers $m$, so we evaluate functions with poles of order at most $m$ at the point at infinity.

\begin{definition}[Hermitian code]
The Hermitian code $\mathcal{C}_t$ is the algebraic geometry code $C_L(\mathcal{H}, \mathcal{P}, t P_\infty)$, where $\mathcal{H}$ is the Hermitian curve and $\mathcal{P}$ consists of all $q^3$ affine rational points.
\end{definition}
The space $L(t P_\infty)$ consists of rational functions that are regular everywhere except possibly at the point at infinity, where they may have poles of order at most $t$. These are essentially bivariate polynomials in the affine coordinates $x$ and $y$ of the Hermitian curve. The parameters of a Hermitian code are:
\begin{itemize}
    \item \textbf{Length}: $n = q^3$
    \item \textbf{Dimension}: $k = t - g + 1 = t - q(q-1)/2 + 1$ (when $t > 2g - 2$)
    \item \textbf{Minimum distance}: $d \geq q^3 - t$
\end{itemize}
For appropriate choices of $t$, these codes achieve excellent rate-distance trade-offs, often approaching the Singleton bound.

\subsubsection{Duality of Hermitian Codes}

A key property for DQI applications is that Hermitian codes have a particularly clean duality structure.

\begin{theorem}[Duality of Hermitian codes]
The dual of the Hermitian code $\mathcal{C}_t$ is isomorphic to the Hermitian code $\mathcal{C}_{t'}$ where
\begin{equation}
    t' = q^3 + 2g - 2 - t = q^3 + q(q-1) - 2 - t.
\end{equation}
\end{theorem}

This means both the primal and dual codes are Hermitian codes on the same curve, ensuring that efficient decoding algorithms exist for both. This duality property is essential for the syndrome decoding step in DQI, which requires efficient decoding of the dual code.

\end{document}